\documentclass{elsarticle}
\usepackage{amsmath,amstext,amsthm,amssymb,epsf}

\newtheorem{theorem}{\sc Theorem}

\newtheorem{coro}{\sc Corollary}
\newtheorem{req}{\sc Requirement}
\newtheorem{nota}{\sc Notation}
\newtheorem{defin}{\sc Definition}
\newtheorem{rem}{\sc Remark}
\newtheorem{cla}{\sc Claim}
\newtheorem{ex}{\sc Example}

\title{Turing Machines and Understanding Computational Complexity}
\author{Paul M.B.Vit\'anyi\\ CWI, Science Park 123, 1098XG Amsterdam\\ The Netherlands}
\date{}
\begin{document}

\begin{abstract}
We describe the Turing Machine, list some of its many influences on the theory
of computation and complexity of computations, and illustrate its importance.
\end{abstract}
\maketitle

\section{Introduction}
A {\em Turing machine} refers to a hypothetical machine proposed by Alan M. Turing (1912--1954) in 1936 \cite{Tu36} whose computations are intended to give an operational and formal definition of the intuitive notion of computability in the discrete domain. It is a digital device and sufficiently simple to be amenable to theoretical analysis and sufficiently powerful to embrace everything in the discrete domain that is intuitively computable. As if that were not enough, in the theory of computation many major complexity classes can be easily characterized by an appropriately restricted Turing machine; notably the important classes P and NP and consequently the major question whether P equals NP.

Turing gave a brilliant demonstration that everything that can be reasonably said to be computed by a human computer using a fixed procedure can be computed by such a machine. As Turing claimed, any process that can be naturally called an effective procedure is realized by a Turing machine. This is known as Turing's thesis. Enter Alonzo Church (1903--1995). Over the years, all serious attempts to give precise yet intuitively satisfactory definitions of a notion of effective procedure (what Church called effectively calculable function) in the widest possible sense have turned out to be equivalent---to define essentially the same class of processes. The Church-Turing thesis states that a function on the positive integers is effectively calculable if and only if it is computable. An informal accumulation of the tradition in S. C. Kleene (1952) \cite{Kl52} has transformed it to the Computability thesis: there is an objective notion of effective computability independent of a particular formalization. The informal arguments Turing sets forth in his 1936 paper are as lucid and convincing now as they were then. To us it seems that it is the best introduction to the subject. It gives the intu\"itions that lead up to the formal definition, and is in a certain sense a prerequisite of what follows. The reader can find this introduction in \cite{Tu36} included in this volume.  It begins with:
\begin{quote}
{\small
``All arguments are bound to be,
fundamentally, appeals to intuition, and for that reason
rather unsatisfactory mathematically. The real question at
issue is: `what are the possible processes which can be
carried out in computing (a number)?'
The arguments which I shall use are of three kinds.
\begin{itemize}
\item[{(a)}]
A direct appeal to intuition.
\item[{(b)}]
A proof of equivalence of two definitions (in case the new
definition has a greater intuitive appeal).
\item[{(c)}]
Giving examples of large classes of numbers which are computable.''
\end{itemize}
}
\end{quote}

\section{Formal Definition of the Turing Machine}

We formalize Turing's description as follows: A Turing machine consists of a finite program, called the finite control, capable of manipulating a linear list of cells, called the tape, using one access pointer, called the head. We refer to the two directions on the tape as {\em right} and {\em left}. The finite control can be in any one of a finite set of states $Q$, and each tape cell can contain a 0, a 1, or a {\em blank} $B$. Time is discrete and the time instants are ordered $0,1,2, \ldots ,$ with 0 the time at which the machine starts its computation. At any time, the head is positioned over a particular cell, which it is said to {\em scan}. At time 0 the head is situated on a distinguished cell on the tape called the {\em start cell}, and the finite control is in a distinguished state $q_0$. At time 0 all cells contain $B$s, except for a contiguous finite sequence of cells, extending from the start cell to the right, which contain 0's and 1's. This binary sequence is called the {\em input}. The device can perform the following basic operations:

\begin{enumerate}
\item  It can write an element from $A= \{0,1,B\}$ in the cell it scans; and
\item it can shift the head one cell left or right.
\end{enumerate}

When the device is active it executes these operations at the rate of one operation per time unit (a {\em step}). At the conclusion of each step, the finite control takes on a state from $Q$. The device is constructed so that it behaves according to a finite list of rules. These rules determine, from the current state of the finite control and the symbol contained in the cell under scan, the operation to be performed next and the state to enter at the end of the next operation execution.

\begin{figure}[t]
    \epsfxsize=3in
    \leftline{\hskip8pc\epsfbox{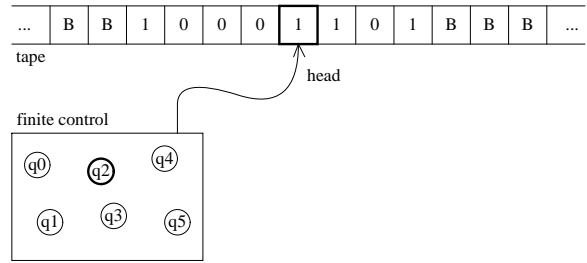}}
    \caption{Turing machine}
\label{fig.turing}
\end{figure}

The rules have format $(p,s,a,q)$: $p$ is the current state of the finite control; $s$ is the symbol under scan; $a$ is the next operation to be executed of type (1) or (2) designated in the obvious sense by an element from $S = \{ 0,1,B,L,R \}$; and $q$ is the state of the finite control to be entered at the end of this step.

For now, we assume that there are no two distinct quadruples that have their first two elements identical, the device is deterministic. Not every possible combination of the first two elements has to be in the set; in this way we permit the device to perform `no' operation. In this case we say that the device halts. Hence, we can define a Turing machine by a mapping from a finite subset of $Q \times A$ into $S \times Q$. Given a Turing machine and an input, the Turing machine carries out a uniquely determined succession of operations, which may or may not terminate in a finite number of steps.

Strings and natural numbers are occasionally identified according to the pairing
\begin{equation}\label{eq.pairing}
    (\epsilon,0), (0,1), (1,2), (00,3), (01,4), (10,5), (11,6), \ldots ,
\end{equation}
where $\epsilon$ denotes the empty string (with no bits). In the following we need the notion of a {\em self-delimiting} code of a binary string. If $x=x_1 \ldots x_n$ is a string of $n$ bits, then its self-delimiting code is $\bar{x}=1^n0x$. Clearly, the length $|\bar{x}| = 2|x|+1$. Encoding a binary string self-delimitingly enables a machine to determine where the string ends reading it from left to right in a single pass and without reading past the last bit of the code.

\subsection{Computable Functions}

We can associate a partial function with each Turing machine in the following way: The input to the Turing machine is presented as an $n$-tuple $(x_1 , \ldots , x_n )$ consisting of self-delimiting versions of the $x_i$'s. The integer represented by the maximal binary string (bordered by blanks) of which some bit is scanned, or 0 if a blank is scanned, by the time the machine halts, is called the {\em output} of the computation. Under this convention for inputs and outputs, each Turing machine defines a partial function from $n$-tuples of integers onto the integers, $n \geq 1$. We call such a function partial computable. If the Turing machine halts for all inputs, then the function computed is defined for all arguments and we call it total computable. (Instead of {\em computable} the more ambiguous {\em recursive} has also been used.) We call a function with range $\{ 0,1 \}$ a {\em predicate}, with the interpretation that the predicate of an $n$-tuple of values is {\em true} if the corresponding function assumes value 1 for that $n$-tuple of values for its arguments and is {\em false} or {\em undefined} otherwise. Hence, we can talk about {\em partial (total) computable predicates}.

\subsection{Examples of Computable Functions}

Consider $x$ as a binary string. It is easy to see that the functions $|x|$, $f(x)= \bar x$, $g( \bar x y)=x$, and $h( \bar x y)=y$ are partial computable. Functions $g$ and $h$ are not total since the value for input 1111 is not defined. The function $g'( \bar x y)$ defined as 1 if $x = y$ and as 0 if $x \neq y$ is a computable predicate. Consider $x$ as an integer. The following functions are basic $n$-place total computable functions: the {\em successor} function $\gamma^{(1)} (x) = x+1$, the {\em zero} function $\zeta^{(n)} (x_1 , \ldots , x_n ) = 0$, and the {\em projection} function $\pi_m^{(n)} (x_1 , \ldots , x_n ) = x_m (1 \leq m \leq n)$. The function $\langle x, y\rangle = \bar x y$ is a total computable one-to-one mapping from pairs of natural numbers into the natural numbers. We can easily extend this scheme to obtain a total computable one-to-one mapping from $k$-tuples of integers into the integers, for each fixed $k$. Define $\langle n_1 , n_2 , \ldots ,n_k \rangle = \langle n_1 , \langle n_2 , \ldots , n_k \rangle \rangle$. Another total recursive one-to-one mapping from $k$-tuples of integers into the integers is $\langle n_1 , n_2 , \ldots ,n_k \rangle = \bar n_1 \ldots \bar n_{k-1} \bar n_k$.

\section{Computability Thesis and the Universal Turing Machine}

The class of algorithmically computable numerical functions (in the intuitive sense) coincides with the class of partial computable functions. Originally intended as a proposal to henceforth supply intuitive terms such as `computable' and `effective procedure' with a precise meaning, the Computability thesis has come into use as shorthand for a claim that from a given description of a procedure in terms of an informal set of instructions we can derive a formal one in terms of Turing machines.

It is possible to give an effective (computable) one-to-one pairing between natural numbers and Turing machines. This is called an {\em effective enumeration}. One way to do this is to encode the table of rules of each Turing machine in binary, in a canonical way.

The only thing we have to do for every Turing machine is to encode the defining mapping $T: Q \times A \rightarrow S \times Q$. Giving each element of $Q \bigcup S$ a unique binary code requires $s$ bits for each such element, with $s = \lceil \log (|Q|+5) \rceil$. Denote the encoding function by $e$. Then the quadruple $(p,0,B,q)$ is encoded as $e(p)e(0)e(B)e(q)$. If the number of rules is $r$, then $r \leq 3|Q|$. We agree to consider the state of the first rule as the start state. The entire list of quadruples,
\[
    T = ( p_1 ,t_1 ,s_1, q_1 ) , (p_2 , t_2 ,s_2 , q_2 ), \ldots , (p_r , t_r , s_r , q_r ),
\]
is encoded as
\[
    E(T) = \bar s \bar r e( p_1 ) e( t_1 ) e( s_1 ) e( q_1 ) \ldots e(p_r ) e( t_r ) e( s_r ) e ( q_r ) .
\]
\noindent
Note that $l(E(T)) \leq 4rs + 2 \log rs + 4$. (Moreover, $E$ is self-delimiting, which is convenient in situations in which we want to recognize the substring $E(T)$ as prefix of a larger string.)

We order the resulting binary strings lexicographically (according to increasing length). We assign an index, or G\"odel number, $n(T)$ to each Turing machine $T$ by defining $n(T)=i$ if $E(T)$ is the $i$th element in the lexicographic order of Turing machine codes. This yields a sequence of Turing machines $T_1 ,T_2 , \ldots$ that constitutes the effective enumeration. One can construct a Turing machine to decide whether a given binary string $x$ encodes a Turing machine, by checking whether it can be decoded according to the scheme above, that the tuple elements belong to $Q \times A \times S \times Q$, followed by a check whether any two different rules start with the same two elements. This observation enables us to construct {\em universal} Turing machines.

A universal Turing machine $U$ is a Turing machine that can imitate the behavior of any other Turing machine $T$. It is a fundamental result that such machines exist and can be constructed effectively. Only a suitable description of $T$'s finite program and input needs to be entered on $U$'s tape initially. To execute the consecutive actions that $T$ would perform on its own tape, $U$ uses $T$'s description to simulate $T$'s actions on a representation of $T$'s tape contents. Such a machine $U$ is also called {\em computation universal}. In fact, there are infinitely many such $U$'s.

We focus on a universal Turing machine $U$ that uses the encoding above. It is not difficult, but tedious, to define a Turing machine in quadruple format that expects inputs of the format $E(T)p$ and is undefined for inputs not of that form. The machine $U$ starts to execute the successive operations of $T$ using p as input and the description $E(T)$ of $T$ it has found so that $U(E(T)p)=T(p)$ for every $T$ and $p$. We omit the explicit construction of $U$.

For the contemporary reader there should be nothing mysterious in the concept of a general-purpose computer which can perform any computation when supplied with an appropriate program. The surprising thing is that a general-purpose computer can be very simple: M. Minsky \cite{Mi67} has shown that four tape symbols and seven states suffice easily in the above scheme. This machine can be changed to, in the sense of being simulated by, our format using tape symbols $\{ 0, 1, B \}$ at the cost of an increase in the number of states. The last reference contains an excellent discussion of Turing machines, their computations, and related machines. The effective enumeration of Turing machines $T_1 ,T_2 , \ldots$ determines an effective enumeration of partial computable functions $\phi_1 , \phi_2 , \ldots$ such that $\phi_i$ is the function computed by $T_i$, for all $i$. It is important to distinguish between a function $\psi$ and a name for $\psi$. A name for $\psi$ can be an algorithm that computes $\psi$, in the form of a Turing machine $T$. It can also be a natural number $i$ such that $\psi$ equals $\phi_i$ in the above list. We call $i$ an index for $\psi$. Thus, each partial computable $\psi$ occurs many times in the given effective enumeration, that is, it has many indices.

\section{Undecidability of the Halting Problem}

Turing's paper \cite{Tu36}, and more so Kurt G\"odel's paper \cite{Go31}, where such a result first appeared, are celebrated for showing that certain well-defined questions in the mathematical domain cannot be settled by any effective procedure for answering questions. The following 'machine form' of this undecidability result is due to Turing and Church: ``which machine computations eventually terminate with a definite result, and which machine computations go on forever without a definite conclusion?'' This is sometimes called the halting problem.

Since all machines can be simulated by the universal Turing machine $U$, this question cannot be decided in the case of the single machine $U$, or more generally for any other individual universal machine. The following theorem due to Turing \cite{Tu36}, formalizes this discussion. Let $\phi_1 , \phi_2 , \ldots$ be the standard enumeration of partial computable functions and write $\phi(x) < \infty$ if $\phi(x)$ is defined and write $\phi(x) = \infty$ otherwise. Define $K_0 = \{ \langle x, y\rangle  : \phi_x (y) < \infty \}$ as the {\em halting set}.

\begin{theorem}
The halting set $K_0$ is not computable.
\end{theorem}

The theorem of Turing on the incomputability of the halting set was preceded by (and was intended as an alternative way to show) the famous (first) incompleteness theorem of Kurt G\"odel in 1931 \cite{Go31}. Recall that a formal theory $T$ consists of a set of well-formed formulas, formulas for short. For convenience these formulas are taken to be finite binary strings. Invariably, the formulas are specified in such a way that an effective procedure exists that decides which strings are formulas and which strings are not.

The formulas are the objects of interest of the theory and constitute the meaningful statements. With each theory we associate a set of true formulas and a set of provable formulas. The set of true formulas is {\em true} according to some (often nonconstructive) criterion of truth. The set of provable formulas is {\em provable} according to some (usually effective) syntactic notion of proof.

A theory $T$ is simply any set of formulas. A theory is axiomatizable if it can be effectively enumerated. For instance, its axioms (initial formulas) can be effectively enumerated and there is an effective procedure that enumerates all proofs for formulas in $T$ from the axioms. A theory is decidable if it is a recursive set. A theory $T$ is consistent if not both formula $x$ and and its negation $\neg x$ are in $T$. A theory $T$ is sound if each formula $x$ in $T$ is true (with respect to the standard model of the natural numbers).

Hence, soundness implies consistency. A particularly important example of an axiomatizable theory is Peano arithmetic, which axiomatizes the standard elementary theory of the natural numbers.

\begin{theorem}
There is a computably enumerable set, say the set $K_0$ defined above, such that for every axiomatizable theory $T$ that is sound and extends Peano arithmetic, there is a number $n$ such that the formula ``$n \not\in K_0$'' is true but not provable in $T$.
\end{theorem}

In his original proof, G\"odel uses diagonalization to prove the incompleteness of any sufficiently rich logical theory $T$ with a computably enumerable axiom system, such as Peano arithmetic. By his technique he exhibits for such a theory an explicit construction of an undecidable statement $y$ that says of itself {\em I am unprovable in $T$}. The formulation in terms of computable function theory is due to A. Church and S.C. Kleene.

Turing's idea was to give a formal meaning to the notion of ``giving a proof.'' Intuitively, a proof is a sort of computation where every step follows (and follows logically) from the previous one, starting from the input. To put everything as broad as possible, Turing analyses the notion of `computation' from an `input' to an `output' and uses this to give an alternative proof of G\"odel's theorem.

Prominent examples of uncomputable functions are the Kolmogorov complexity function and the universal algorithmic probability function. These are the fundamental notions in \cite{LV08} and, among others, \cite{Ni09,DH10}.

\section{Complexity of Computations}

Theoretically, every intuitively computable (effectively calculable) function is computable by a personal computer or by a Turing machine. But a computation that takes $2^n$ steps on an input of length $n$ would not be regarded as practical or feasible. No computer would ever finish such a computation in the lifetime of the universe even with $n$ merely $1,000$. For example, if we have $10^9$ processors each taking $10^9$ steps/second, then we can execute $3.1 \times 10^{25} < 2^{100}$ steps/year. Computational complexity theory tries to identify problems that are feasibly computable.

In computational complexity theory, we are often concerned with languages. A language $L$ over a finite alphabet $\Sigma$ is simply a subset of $\Sigma^*$. We say that a Turing machine accepts a language $L$ if it outputs 1 when the input is a member of $L$ and outputs 0 otherwise. That is, the Turing machine computes a predicate.

Let $T$ be a Turing machine. If for every input of length $n$ we have that $T$ makes at most $t(n)$ moves before it halts, then we say that $T$ runs in time $t(n)$, or has time complexity $t(n)$. If $T$ uses at most $s(n)$ tape cells in the above computations, then we say that $T$ uses $s(n)$ space, or has space complexity $s(n)$.

For convenience, we often give the Turing machine in Figure~\ref{fig.turing} a few more work tapes and designate one tape as a read-only input tape. Thus, each transition rule will be of the form $(p,{\bar s},a,q)$, where $\bar s$ contains the scanned symbols on all the tapes, and $p,a,q$ are as above, except that an operation now involves moving maybe more than one head.

We sometimes also make a Turing machine nondeterministic by allowing two distinct transition rules to have identical first two components. That is, a nondeterministic Turing machine may have different alternative moves at each step. Such a machine accepts if one accepting path leads to acceptance. Turing machines are deterministic unless it is explicitly stated otherwise.

\begin{itemize}
\item     ${\rm DTIME}[t(n)]$ is the set
of languages accepted by multitape deterministic Turing machines in time $O(t(n))$;
\item
     ${\rm NTIME}[t(n)]$ is the set of languages accepted by multitape
nondeterministic Turing machines in time $O(t(n))$;
\item
     ${\rm DSPACE}[s(n)]$ is the set of languages accepted by multitape
deterministic Turing machines in $O(s(n))$ space;
\item
     ${\rm NSPACE}[s(n)]$ is the set of languages accepted by multitape
nondeterministic Turing machines in $O(s(n))$ space. 
\item
With $c$ running through the natural numbers:

     P is the complexity class $\bigcup_{c } {\rm DTIME}[n^c]$;

     NP is the complexity class $\bigcup_{c } {\rm NTIME}[n^c]$;

     PSPACE is the complexity class $\bigcup_{c } {\rm DSPACE}[n^c]$.
\end{itemize}

Languages in P, that is, languages acceptable in polynomial time, are considered feasibly computable. The nondeterministic version for PSPACE turns out to be identical to PSPACE by Savitch's Theorem \cite{Sa70} which states that ${\rm NSPACE}[s(n)]={\rm DSPACE}[(s(n))^2]$. The following relationships hold trivially, ${\rm P} \subseteq {\rm NP} \subseteq {\rm PSPACE}$. It is one of the most fundamental open questions in computer science and mathematics to prove whether either of the above inclusions is proper. Research in computational complexity theory focuses on these questions. In order to solve these problems, one can identify the hardest problems in NP or PSPACE.
The Bible of this area is \cite{GJ}.

\section{Importance of the Turing Machine}

In the last three-quarter of a century the Turing machine model has proven to be of priceless value for the development of the science of data processing. All theory development reaches back to this format. The model has become so dominant that new other models that are not polynomial-time reducible to Turing machines are viewed as not realistic (the so-called polynomial-time Computability thesis). 

Without explaining terms, the {\em random access machine} (RAM) with {\em logarithmic cost}, or {\em unit cost} without multiplications, is viewed as realistic, while the unit cost RAM with multiplications or the {\em parallel random access machine} (PRAM) are not so viewed. New notions, such as randomized computations as in R. Motwani and P. Raghavan \cite{MR} (like the fast primality tests used in Internet cryptographical protocols) are analysed using {\em probabilistic} Turing machines. 

In 1980 the Nobelist Richard Feynman proposed a {\em quantum computer}, in effect an analogous version of a quantum system. Contrary to digital computers (classical, quantum, or otherwise), an analogue computer works with continuous variables and simulates the system we want to solve directly: for example, a wind tunnel with a model aircraft simulates the aeroflow and in particular nonlaminar turbulence of the aimed-for actual aircraft. In practice, analogue computers have worked only for special problems. In contrast, the digital computer, where everything is expressed in bits, has proven to be universally applicable. Feynman's innovative idea was without issue until D. Deutsch \cite{De} put the proposal in the form of a quantum Turing machine, that is, a digital quantum computer. This digital development exploded the area both theoretically and applied to the great area of {\em quantum computing}.


\begin{thebibliography}{99}

\bibitem{De}
     D. Deutsch, Quantum theory, the Church-Turing principle and the universal quantum computer, {\em Proc. Royal Society of London} A, 400(1985), 97--117.

\bibitem{DH10}
R.G. Downey, D. Hirschfeldt, {\em Algorithmic Randomness and Complexity (Theory and Applications of Computability)}, Springer, New York, 2010.

\bibitem{GJ}
     M.R. Garey and D.S. Johnson, {\em Computers and Intractability: A Guide to the Theory of {\rm NP}-Completeness}, W.H. Freeman, 1979.

\bibitem{Go31}
     K. G\"odel, \"Uber formal unentscheidbare Sätze der Principia Mathematica und verwandter Systeme, I, {\em Monatshefte f\"ur Mathematik und Physik}, 38(1931), 173--198.

\bibitem{Kl52}
     S. C. Kleene, {\em Introduction to Metamathematics}, Van Nostrand, New York, 1952.

\bibitem{LV08}
     M. Li, P.M.B. Vit\'anyi, {\em An Introduction to Kolmogorov Complexity and Its Applications}, Springer, New York, Third edition, 2008.

\bibitem{Mi67}
     M. Minsky, {\em Computation: Finite and Infinite Machines}, Prentice-Hall, Inc. Englewood Cliffs, NJ, 1967.

\bibitem{MR}
     R. Motwani and P. Raghavan, {\em Randomized Algorithms}, Cambridge Univ. Press, 1995.


\bibitem{Ni09}
     A. Nies, {\em Computability and Randomness}, Oxford Univ. Press, USA, 2009.

\bibitem{Sa70}
W.J. Savitch, Relationships between nondeterministic and deterministic tape complexities, {\em J. Computer and System Sciences}, 4:2(1970), 177--192


\bibitem{Tu36}
     A.M. Turing, On Computable Numbers, with an Application to the Entscheidungsproblem, {\em Proc. London Mathematical Society} 2, 42(1936), 230--265, "Correction", 43(1937), 544--546.
\end{thebibliography}
\end{document}